\documentclass[aps,floatfix,nofootinbib,
preprint]{revtex4}

\usepackage{graphicx}
\usepackage{epic}
\usepackage{eepic}
\usepackage{latexsym}

\newcommand{\eq}[1]{(\ref{#1})}
\newcommand{\be}{\begin{equation}}
\newcommand{\ee}{\end{equation}}
\newcommand{\bea}{\begin{eqnarray}}
\newcommand{\eea}{\end{eqnarray}}

\newcommand{\vs}[1]{\vspace{#1 mm}}
\newcommand{\hs}[1]{\hspace{#1 mm}}

\def\d{\delta}

\def\fr{\frac}

\def\l{\lambda}

\def\m{\mu}
\def\n{\nu}

\def\r{\rho}

\def\O{\Omega}

\def\o{\omega}

\let\bm=\bibitem
\def\nn{\nonumber}

\begin{document}

\title{Cosmological Evolution of Vacuum and Cosmic Acceleration}

\author{Ali Kaya}
\email[]{ali.kaya@boun.edu.tr}
\affiliation{\large Bo\~{g}azi\c{c}i University, Department of Physics, \\ 34342,
Bebek, \.{I}stanbul, Turkey \vs{9}}

\begin{abstract}

\vs{9}
It is known that the unregularized expressions for the stress-energy tensor 
components corresponding to  subhorizon and  superhorizon vacuum
fluctuations of a massless scalar field in a Friedmann-Robertson-Walker 
background are characterized by the equation of state parameters $\o=1/3$ and
$\o=-1/3$, which are not sufficient to produce cosmological acceleration. 
However, the form of the adiabatically regularized finite stress-energy tensor
turns out to be completely different. By using the fact that vacuum subhorizon
modes evolve nearly adiabatically and superhorizon modes have $\o=-1/3$, we 
approximately determine the regularized stress-energy tensor, whose
conservation is utilized to fix the time dependence of the vacuum energy
density. We then show that vacuum energy density grows from zero up to $H^4$ in about 
one Hubble time, vacuum fluctuations give positive acceleration of the
order of $H^4/M_p^2$ and they can completely alter the 
 cosmic evolution of the universe  dominated otherwise by cosmological constant, radiation
 or pressureless dust. Although the magnitude of the acceleration
is tiny  to explain the observed value today, our findings indicate that the
cosmological backreaction of vacuum fluctuations must be taken into account in
early stages of cosmic evolution.  
\end{abstract}

\maketitle

\section{Introduction}

Explaining the observed cosmological acceleration of the universe is one of the
most important and difficult problems of modern cosmology awaiting a solution.
Whether the acceleration is due to a bare  cosmological constant or mysterious
dark energy, understanding its cause in terms of fundamental physics is
challenging. This leads many researchers to seek for unorthodox solutions and
there are a lot of new ideas and directions in the literature claiming different
answers to the problem. In the absence of any other experimental or
observational input to test these ideas, it is difficult to judge them based
solely on theoretical arguments and decide whether a natural solution is really
obtained.  

On the other hand, one may wonder if at least a partial explanation in terms of
standard physical notions can be given to the acceleration phenomena. In this
article, we show by a careful  treatment that  quantum vacuum fluctuations of a
massless scalar field in a Friedmann-Robertson-Walker (FRW) space-time can give
cosmological acceleration. Usually, vacuum fluctuations of quantized fields are thought to
 contribute to the bare cosmological constant. In fact, the famous statement 
that there is a $10^{122}$  order of magnitude discrepancy  
between the expected and the observed values of
the cosmological constant depends on this view (see e.g. \cite{w}). However,
the cosmological constant is characterized by the equation of state $P=\o\rho$ with $\o=-1$,
which cannot easily be fulfilled by vacuum fluctuations. For example, 
the (unregularized) stress-energy tensor of a quantized massless scalar field in
Minkowski space-time has $\o=1/3$, which is naively equivalent to the radiation.
Similarly, for a massless scalar in an expanding FRW universe while the
subhorizon modes are described by the parameter $\o=1/3$, the superhorizon modes
have $\o=-1/3$, which are not close to the equation of state of the
cosmological constant. As discussed in \cite{tp1,tp2} it may not even be
possible to mimic cosmological constant by quantum fluctuations
with trans-Planckian modifications of the dispersion relation.  

One may then think that the quantum vacuum fluctuations cannot give rise to
cosmic acceleration, which requires $\o<-1/3$. However, the above mentioned
naive expressions for the stress-energy tensor components contain ultraviolet 
(UV) divergences and they must be regularized in a suitable way, which may alter
the equation of state (this will be the case as we will see below). Moreover,
 vacuum fluctuations and the Hubble parameter influence each other in a very
non-trivial way: while the expansion speed separates the subhorizon and
superhorizon excitations, the vacuum energy density directly affects the
magnitude of the Hubble parameter. In other words, simple  general arguments
based solely on the equation of state might be misleading since the dynamical
impact of the vacuum  fluctutaions is different than that of a single
component perfect fluid.

In this article, we study the cosmological evolution of a quantized massless
scalar field in a FRW space-time placed initially in its ground state. 
The problem is truly very hard to tackle:  the adiabatically regularized
finite stress-energy tensor, which drives the expansion, can be determined in
terms of the integrals of the mode functions, which in turn obey the massless
Klein-Gordon equation on the background. Therefore, one encounters a
complicated but consistent system of  coupled integro-differential equations
involving the scale factor of the universe and the mode functions. The assumption that
the scalar is in its ground state at some time specifies the initial conditions
for the mode functions. Supplying also the expansion speed of the universe at
the same time as an initial condition, the system of equations can in principle
be uniquely solved to determine the subsequent evolution. 

Of course it is very difficult, if not impossible, to exactly solve this
complicated system. In order to figure out the cosmological evolution, we try to
simplify these equations by using the fact that the subhorizon
modes must evolve adiabatically and the superhorizon modes can be characterized
by the equation of state parameter $\o=-1/3$. In this way, the stress-energy
tensor can approximately be found without knowing the time dependence of
the mode functions explicitly. Instead, the vacuum energy density can be shown
to obey a modified conservation equation determining its time evolution. The
final, approximate but consistent differential equation system is simple enough
to analyze analytically.

Partially due to the terms coming from adiabatic regularization, the
stress-energy tensor of vacuum fluctuations turns out to depend on the Hubble
parameter and its time derivative. Consequently, the cosmological 
evolution is significantly altered and one can see from the field equations 
that the expansion is always accelerated whose order of magnitude is given by 
$H^4/M_p^2$.  As a matter of fact, it is possible to obtain an analytical
solution describing  the cosmological evolution of vacuum.  Depending on the initial
expansion speed, the solution gives an increasing Hubble parameter blowing up in
a finite proper time (i.e. a big-rip singularity) or a constant Hubble parameter
(i.e. de Sitter space) or a decreasing but always positive acceleration
parameter which asymptotically vanishes.  
Although the magnitude of the acceleration $H^4/M_p^2$ is very small to
explain the accelerated expansion today, it is encouraging to see that a
seemingly simple and standard physical system yields cosmic acceleration.
Besides, we observe that vacuum fluctuations can modify the standard cosmic 
evolution of matter in a very non-trivial way. For example, in the presence of a
cosmological constant we show that the Hubble parameter increases in time and
it either blows up in a finite proper time or asymptotes to a larger fixed value. 
Similarly, when the initial expansion speed is about the Planck
scale, the radiation or dust dominated expansions are also significantly
modified. While the validity of this last statement is questionable due to the possible 
modifications of the standard
equations near Planck scale, it still indicates that a proper understanding of
vacuum dynamics in an expanding  universe is very crucial in early time
cosmology. 

Without a doubt, the problem of determining the stress-energy tensor of a quantized
field in a cosmological background is a well studied problem. For example, it was
shown in an earlier work \cite{pf} that a massive quantized field in a closed FRW geometry can
stop a cosmological collapse and convert it to an expansion, hence avoiding a
cosmic singularity. Similarly, it was pointed out in \cite{d} that due to the
trace anomaly induced by the broken conformal invariance, the strong energy
condition can be violated, which indeed gives acceleration. Although our approach in this paper is akin to that of \cite{pf,d}, the final set of equations we obtain are much simpler and easy to interpret physically.  
Consequently, we are able to construct analytical solutions describing the  cosmological impact of vacuum. On the other hand, more recently, the backreaction of a quantum scalar field in a curved spacetime has also been analyzed in relation to the cosmic acceleration phenomena (see e.g. \cite{pr1,pr2,pr3,new1,new2,deb1,deb2}). In the next section, we will compare our findings with these results already existed in the literature. 

\section{Cosmological evolution of vacuum}

We consider a massless scalar field $\phi$ in a FRW space-time which has the
metric 
\bea
ds^2&=&-dt^2+a(t)^2(dx^2+dy^2+dz^2)\nn\\
&=&a(\eta)^2(-d\eta^2+dx^2+dy^2+dz^2).
\eea
The corresponding Hubble parameters in the conformal and the proper time
coordinates are defined as 
\be 
h=\fr{a'}{a},\hs{10}H=\fr{\dot{a}}{a},
\ee
where the prime and the dot represent derivatives with respect to $\eta$ and $t$,
respectively. In this paper we only consider expanding geometries and thus assume $H>0$. 
A real, massless scalar field $\phi$ propagating in this
background obeys 
\be
\nabla^2\phi=0.
\ee
For quantization it is convenient to define a new field $\m$ by
\be
\m=a \,\phi.
\ee
After applying the standard canonical quantization procedure, one can see that
the field operator $\m$ can be decomposed in terms of the {\it time-independent}
ladder operators as 
\be
\m=\int \fr{d^3
k}{(2\pi)^{3/2}}\left[\m_k(\eta)\,e^{i\vec{k}.\vec{x}}\,a_{\vec{k}}
+\m_k(\eta)^*\,e^{-i\vec{k}.\vec{x}}\,a_{\vec{k}}^\dagger\right], 
\ee
where $\vec{k}$ is the comoving momentum variable,
$[a_{\vec{k}},a^\dagger_{\vec{k}'}]=\d(\vec{k}-\vec{k}')$ and  the mode
functions satisfy  the Wronskian condition $\m_k\m_k'^*-\m_k^*\m_k'=i$ together with 
\be
\m_k''+\left[k^2-\fr{a''}{a}\right]\m_k=0.\label{mf}
\ee
The ground state $|0>$ of the system at time $\eta_0$ can be defined by imposing 
\be
a_{\vec{k}}\,|0>=0,
\ee
and 
\be
\m_k(\eta_0)=\fr{1}{\sqrt{2k}},\hs{10}\m_k'(\eta_0)=-i\sqrt{\fr{k}{2}}. \label{ic}
\ee
The subhorizon and superhorizon modes are given by  
$k>h$ and $k<h$, respectively. The approximate solutions of \eq{mf} in these two regimes
can be determined  as $\m_k\simeq e^{ ik\eta}$ and $(\m_k/a)'\simeq 0$. 

Using the definition of the stress-energy-momentum tensor
$T_{\m\n}=\nabla_\m\phi \nabla_\n\phi-\fr{1}{2}g_{\m\n}(\nabla\phi)^2$, one can
easily calculate the following vacuum-expectation values 
\bea
&&<0|\r|0>=\frac{1}{4\pi^2 a^4}\int_0^\infty k^2\left[ a^2
\left|\left(\fr{\m_k}{a}\right)'\right|^2+k^2|\m_k|^2\right]dk,\nn\\ 
&&<0|P|0>=\frac{1}{4\pi^2 a^4}\int_0^\infty k^2\left[ a^2
\left|\left(\fr{\m_k}{a}\right)'\right|^2-\fr{k^2}{3}|\m_k|^2\right]dk.
\label{se} 
\eea
From these expressions we see that for subhorizon modes obeying $\m_k\simeq
e^{ik\eta}$ (and also for $k\gg h$) the stress-energy tensor is characterized by the equation of state
parameter $\o=1/3$. Likewise, the stress-energy tensor of  superhorizon modes satisfying 
$(\m_k/a)'\simeq 0$ has $\o=-1/3$. One can verify that the conservation equation
$\rho'+3h(\rho+P)=0$ is identically satisfied provided that the mode functions
obey \eq{mf}.  

The expectation values given in \eq{se} contain UV divergences and must be
regularized. The most straightforward way of regularization is to place
a UV cutoff for momentum integrals. However, to preserve stress-energy
conservation the {\it comoving} cutoff scale must be chosen to be time
independent. Thus, the physical cutoff becomes time-dependent which is 
puzzling.  

An alternative way is to apply adiabatic (or WKB) regularization \cite{ad1,ad2}, which
gives finite results without the need of introducing a new scale to the problem.
Let us briefly review the adiabatic regularization procedure. One first writes
the mode function $\m_k$ in terms of a new variable $\O_k$  in the following WKB form  
\be
\m_k=\fr{1}{\sqrt{2\O_k}}\,e^{-i\int \O_k d\eta}. \label{admf} 
\ee
Using \eq{mf}, $\O_k$ can be seen to obey  
\be\label{11}
\O_k^2=k^2-\fr{a''}{a}+\fr{3}{4}\fr{\O_k'^2}{\O_k^2}-\fr{1}{2}\fr{\O_k''}{\O_k}.
\ee
It is now possible to solve this equation iteratively by including
terms with more and more time derivatives (the zeroth order solution is
$\O_k=k$). To regularize \eq{se}, one simply subtracts the corresponding stress-energy tensor
expressions obtained from the adiabatic mode function \eq{admf}. To cancel the quartic and the quadratic divergences it is enough to proceed up to  the adiabatic order two, i.e. to keep the terms containing up to two time derivatives.  To remove the remaining logarithmic divergence, one should also subtract the fourth order adiabatic terms. 

In this paper, we only utilize adiabatic regularization up to second order. As we will see below, after applying a further approximation, one can still obtain a finite and consistent set of field equations even at this order. Furthermore, these equations turn out to be simple enough to be analyzed analytically. When the fourth order adiabatic subtractions are included, the system becomes much more difficult to be examined analytically. Moreover, in that case the scale factor obeys a {\it fourth} order differential equation, which is problematic  when one tries to interpret the dynamical evolution in the Hamiltonian formulation (in that case to determine the evolution uniquely one should know not only the initial expansion speed but also the initial acceleration and the third derivative of the scale factor.)
To avoid the logarithmic divergence,  the momentum integrals may still be imagined to be restricted by the Planck scale. As we will see in a moment, the dependence of the stress-energy tensor on the cut off scale will disappear in the final set of equations. 

Up to adiabatic order two, \eq{11} can be solved as 
\be
\O_k=k\left[1-\fr{a''}{2ak^2}\right].\label{ad2}
\ee
After a straightforward calculation one can then obtain the following regularized expressions for the stress-energy tensor components:  
\bea
&&\r_V=\frac{1}{4\pi^2 a^4}\int_0^\infty k^2\left[ a^2
\left|\left(\fr{\m_k}{a}\right)'\right|^2+k^2|\m_k|^2-k-\fr{h^2}{2k}\right]dk,
\label{frho}\\ 
&&P_V=\frac{1}{4\pi^2 a^4}\int_0^\infty k^2\left[ a^2
\left|\left(\fr{\m_k}{a}\right)'\right|^2-\fr{k^2}{3}|\m_k|^2-\fr{k}{3}-\fr{h^2}
{2k}+\fr{a''}{3ak}\right]dk.\label{fp}
\eea
As shown in \cite{ad1}, $\r_V$ and $P_V$ are guaranteed to be free of quartic and quadratic infinities  and obey the conservation equation $\rho_V'+3h(\rho_V+P_V)=0$. Using them as sources in
the field equations one finds  
\bea
&& h^2=\fr{8\pi a^2}{3M_p^2}\,\r_V,\label{e1}\\
&& h'=-\fr{4\pi a^2}{3M_p^2}\,(\r_V+3P_V).\label{e2}
\eea
Giving  $h(\eta_0)$, setting $a(\eta_0)=1$ and  imposing
further the initial conditions\footnote{The initial conditions defining the
vacuum state must be modified due to the counter-terms added to the action.
Moreover, imposing \eq{ic} for superhorizon modes can be criticized based on the
principle of locality. Since we will not use \eq{ic} in our calculations, we do
not discuss this issue further.}  \eq{ic}, the equations \eq{mf}, \eq{frho}, 
\eq{fp}, \eq{e1} and \eq{e2}  give a complicated but (up to a possibly logarithmic divergence)
well defined  integro-differential equation system for $\m_k$ and the scale factor of the
universe $a(\eta)$, which uniquely determine the cosmic evolution.

This complicated system is very difficult to analyze analytically. To simplify
these equations we point out that there should not be too much difference
between the physical and adiabatic modes for subhorizon excitations with $k>h$.
In other words, subhorizon modes can approximately be viewed to evolve
adiabatically because "the particle creation effects" can be ignored for them.
Therefore, in \eq{frho} and \eq{fp}  the momentum integrals  can be neglected in
the range  $(h,\infty)$, since the terms coming from the mode functions $\m_k$
will be canceled by adiabatic subtractions. On the other hand, by defining  
\bea
&&\r_S=\frac{1}{4\pi^2 a^4}\int_0^h k^2\left[ a^2
\left|\left(\fr{\m_k}{a}\right)'\right|^2+k^2|\m_k|^2\right]dk,
\label{rs}\\ 
&&P_S=\frac{1}{4\pi^2 a^4}\int_0^h k^2\left[ a^2
\left|\left(\fr{\m_k}{a}\right)'\right|^2-\fr{k^2}{3}|\m_k|^2 \right]dk,\label{rp}
\eea
which encode the contributions coming from  the non-adiabatically evolving 
superhorizon modes, one can write  
\bea
&&\r_V=\r_S-\fr{h^4}{8\pi^2 a^4},\label{v1}\\
&&P_V=P_S-\fr{h^4}{12\pi^2 a^4}+\fr{h^2}{24\pi^2 a^4}\fr{a''}{a}.\label{v2}
\eea
One must now fix the time evolution of the superhorizon modes or correspondingly
$\rho_S$ and $P_S$. To achieve this we recall that superhorizon modes
obey $(\m_k/a)'\simeq 0$ and  $P_S=-\r_S/3$. As a result, the
stress-energy conservation can now be used to determine $\r_S'$. Namely, the
conservation equation $\rho_V'+3h(\rho_V+P_V)=0$, which is guaranteed to be
satisfied in the exact system, can be utilized to solve for  the unknown
function $\r_S$.  In this way, the final set of equations in the proper time
coordinate $t$ can be written as 
\bea
&& H^2=\fr{8\pi}{3M_p^2}\,\r_V,\label{f1}\\
&& \fr{\ddot{a}}{a}\equiv
\dot{H}+H^2=\left(1+\fr{H^2}{6\pi M_p^2}\right)^{-1}\,\fr{H^4}{3\pi M_p^2},\label{f2} 
\eea
where $\r_V$ obeys 
\be
\dot{\r}_V+2H\r_V=\fr{1}{8\pi^2}H^5-\fr{1}{8\pi^2}H^3\dot{H}.\label{fc}
\ee
Note that \eq{fc} follows from the conservation equation, which dictates how
vacuum energy density is forced by the expansion of the background. Likewise,   
the vacuum stress-energy tensor alters the standard equation for the scale factor in a 
very non-trivial way and yields \eq{f2}. 
It is straightforward to check that these last three equations
for two unknown functions $H$ and $\r_V$ are consistent, i.e. \eq{f2} follows
from \eq{f1} and \eq{fc}, as it should be. Remarkably, one sees from \eq{f2}
that vacuum fluctuations give rise to acceleration of the order of
$H^4/M_p^2$.  

In fact, there is an important {\it loophole} in the above derivation: in
neglecting the momentum integrals in \eq{frho} and \eq{fp} in the interval
$(h,\infty)$, one assumes that all subhorizon modes with $k>h$ at a given time $t>t_0$ 
have evolved adiabatically in their entire history. This is clearly
wrong for the modes which were born as superhorizon and then later become
subhorizon, since in that case the difference between the real mode $\m_k$ and
the corresponding adiabatic mode cannot be ignored. Therefore, the decomposition
of the stress-energy tensor components given in \eq{v1} and \eq{v2}, and
consequently the field equations \eq{f1}, \eq{f2} and \eq{fc},  can only be used
{\it when the acceleration is always positive in the whole cosmic history},
which forbids the superhorizon modes to turn into subhorizon regime.  From
\eq{f2}, we see that a cosmology dominated by vacuum fluctuations has always
positive acceleration, which is an important self-consistency check of these
equations.  

The above comments also explain why it is not correct to evaluate the integrals in
\eq{frho} and \eq{fp} by using the approximate solutions $\m_k\simeq e^{ik\eta}$
for subhorizon modes and $(\m_k/a)'\simeq 0$ for superhorizon modes. 
In other words, a mode may evolve
from one regime to the other,  and indeed the main reason for the emergence of
acceleration in this setup, despite the fact that neither superhorizon nor
subhorizon modes give $\o<-1/3$, is precisely these crossing overs. We should
note that the terms coming from adiabatic regularization also play a role in obtaining the
acceleration. We thus observe that as regularization changes the naive spectrum of cosmological
perturbations produced during inflation \cite{ad3,ad4},  it also significantly
changes the cosmic evolution due to vacuum fluctuations.    

Naturally one should impose  $\r_V(t_0)=0$  as an initial
condition.\footnote{This actually follows from \eq{frho} if one uses the initial
conditions \eq{ic} in the integral. However, to determine the initial value of
the pressure $P_V(t_0)$ from \eq{fp} is more subtle. In our approach  $P_V(t_0)$
is fixed by the  conservation equation.} However from the Friedmann equation
\eq{f1}, this implies $H(t_0)=0$. The unique solution with these initial
conditions is $H=0$ and $\r_V=0$, which in some sense shows the stability of the flat
space\footnote{In flat space the adiabatic regularization is equivalent to 
normal ordering and thus both vacuum energy density and pressure vanish.}.
To obtain non-trivial cosmological solutions in this picture we assume
that $\r_V(t_0)\not = 0$. Physically, this corresponds to a situation where the
vacuum fluctuations have been created previously and started to  dominate  the
expansion after some time. As we will see in the next section, by adding matter
in the form of cosmological constant, radiation or dust, this scenario can
actually be realized.  

\begin{figure}[htbp]
\begin{center}
\includegraphics[scale=0.69]{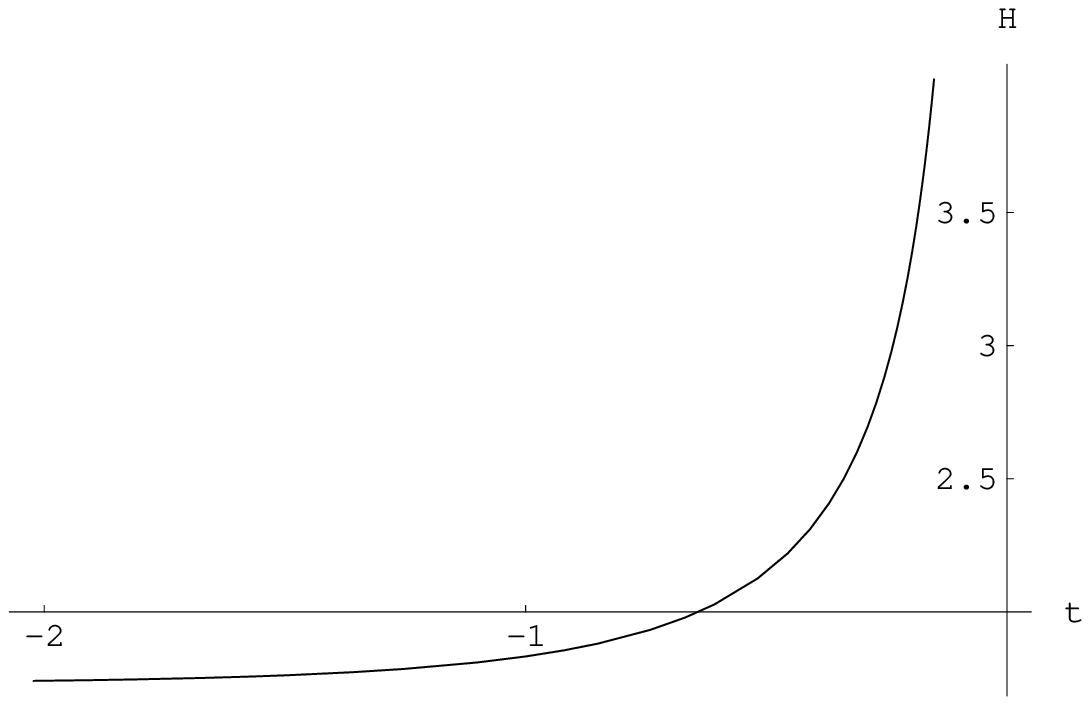}
\includegraphics[scale=0.69]{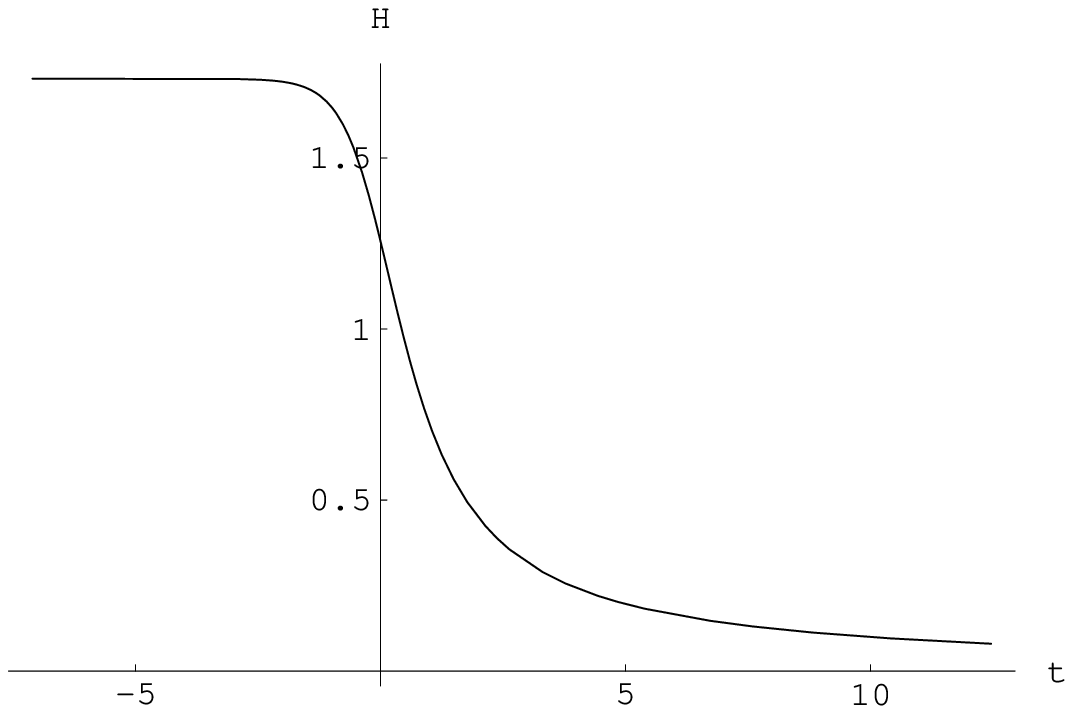}
\vs{3}
(a)\hs{80} (b)
\vs{3}
\caption{The graphs of the  Hubble parameter: (a) for $H>\sqrt{6\pi}M_p$ and (b)
for  $H<\sqrt{6\pi}M_p$, in units $M_p=1$. } 
\label{fig1}
\end{center}
\end{figure}

It is possible to solve \eq{f2} exactly. There is a special solution with
constant Hubble parameter where  
\be
H=\sqrt{6\pi}M_p.
\ee
Other than $H=0$, this is the unique solution with constant $H$ which  mimics
the cosmological constant. For $H\not = \sqrt{6\pi}M_p$, the following implicit
solution can be found 
\be\label{im}
\fr{1}{H}+\fr{1}{\sqrt{6\pi}M_p}\ln\left|\fr{H-\sqrt{6\pi}M_p}{H+\sqrt{6\pi}M_p}\right|=t.
\ee
If $H>\sqrt{6\pi}M_p$, the Hubble parameter can be seen to increase indefinitely
and one encounters a big-rip singularity at finite proper time where $H$ blows
up. If $H<\sqrt{6\pi}M_p$, the Hubble parameter continuously  decreases in time.
Although in that case $\dot{H}<0$, the acceleration is always positive
$\ddot{a}>0$ and one can see that $a\to t$ asymptotically as $t\to\infty$. The
graphs of $H$ for these cases are given in figure \ref{fig1}.  

The non-homogenous equation of motion \eq{fc} shows that in one Hubble time the
vacuum energy density $\r_V$ is driven  by the expansion to increase up to 
$H^4$. However, the homogenous piece of  $\r_V$ decreases
as $\r_V\sim 1/a^2$. Therefore when $H^4$ decreases faster than $1/a^2$ during
cosmological evolution,  $\r_V$ is expected to increase from zero to a maximum value and then 
decrease like $1/a^2$.

One may wonder if quantum vacuum fluctuations can be used 
to realize an inflationary scenario. We found 
that for $H=\sqrt{6\pi}M_p$, the solution is exactly the de Sitter
space, which would give eternal inflation. Alternatively, by tuning $H$
close to this value, it is possible to get a nearly exponential expansion with a
suitable number of e-foldings. Note that exit from inflation is naturally
achieved since $H$ decreases in time.  

We close this section by comparing our results with other approaches in the literature. In \cite{pr1,pr2,pr3}, the cosmological impact of a very light {\it massive} scalar field is studied using  the one-loop effective action, which is determined by applying the zeta-function regularization and heat kernel methods. As shown in \cite{pr1},  the effective action gives the standard trace anomaly in the massless conformally coupled limit. Using the modified field equations, and approximately treating the scalar curvature as a constant, the authors of  \cite{pr1,pr2,pr3} manage to obtain an approximate accelerating solution, which can be glued with continuous first and second derivatives to a matter dominated FRW spacetime. Moreover the free parameters of this solution, which are the mass of the scalar and  its curvature coupling, can be fit to agree with the type Ia supernova data. It is clear that the stress-energy-momentum tensor obtained from the one loop effective action is expected to agree (or at least imply the same physics) with the adiabatically regularized stress-energy tensor. The fact that the trace anomaly of the massless conformally coupled scalar can be obtained from the one loop effective action supports this claim. However, the stress-energy-momentum tensor, either adiabatically regularized or obtained from the one-loop effective action, is very complicated and highly non-linear, involving terms up to four derivatives of the metric functions.  Therefore, it is difficult to trust any approximate background since the non-linear differential equations containing higher derivative terms can easily give run away solutions and stability can be an issue. Moreover,  quantum effects are expected to be much more stronger near big-bang and therefore one should also make sure that the one loop effective action  does not alter the standard cosmological evolution (this is a much more difficult problem to deal with since the curvatures are large). By employing the adiabatic regularization up to second order and canceling the remaining logarithmic divergence by a physical argument, we manage to obtain a very simple set of self consistent field equations, whose solutions physically make sense.  For example, the vacuum energy density obtained from the modified conservation equation \eq{fc} turns out to be positive and grows up to $H^4$, which is consistent with the order of magnitude estimate given in \cite{d}.  An interesting possibility, which is worth to elaborate, is to see whether the one loop effective action of \cite{pr1,pr2,pr3}  truncated in a certain limit  agrees with  the simplified equations obtained in our paper. 

In another interesting work \cite{new1}, the dynamical evolution of a massive quantum scalar field in a FRW background is studied using the Born-Oppenheimer reduction of the mini-superspace Wheeler-De Witt equation. In this way, the authors of \cite{new1} succeed to obtain coupled equations which describe the backreaction of quantum matter on the semiclassical evolution of the  scale factor of the universe obeying a Hamilton-Jacobi type equation. They show that an inflationary period is possible for a large set of initial quantum states. The approach of \cite{new1} mainly differs from ours in that the scale factor is treated semiclassically using the mini-superspace approximation (in our case the background metric is assumed to be purely classical).  Therefore, in some way they accomplish to take into account the "quantum gravitational effects". On the other hand, the matter sector is analyzed not in the field theory context, but in its quantum mechanical truncation, which can be seen as a drawback of this method. 

To realize an inflationary scenario based on quantum backreaction effects, rather than using classical scalar fields,  in \cite{new2} a class of purely gravitational but non-local models are studied. Quantum gravitational corrections are encoded in an effective energy momentum tensor, which is assumed to have the perfect fluid form. These models are characterized by a distinctive phase of oscillations, which occur both during and after inflation. The oscillations terminate at the onset of matter domination and  can induce a positive cosmological constant  having the right size. The approach of \cite{new2} is completely different than the method of the present paper. Whereas in \cite{new2} the contribution of quantum matter is ignored, here we simply neglect the quantum gravitational effects. We can justify our approximation by referring to the main stream view that the quantum gravitational effects must be  negligible on scales smaller than the Planck scale (indeed all inflationary models,  which use classical scalar fields, rely on this common belief), but the findings of \cite{new2} show that in some models this may not be the case. 

Finally, we should mention that,  as discussed in \cite{deb1,deb2}, determining the strength of the quantum backreaction effects in {\it non-linear} theories can be a delicate issue. In \cite{deb1}, the backreaction of the quadratic scalar fluctuations on the leading order linearized metric perturbations is studied in the exact de Sitter spacetime, where the cosmological constant is generated by the background scalar field placed in the minimum of its potential. In general, the linearized equations are shown to have instabilities, which can be avoided by taking strictly de Sitter invariant quantum states.  It is difficult to apply the results of \cite{deb1} directly to our case, since other than the special finely tuned solution with constant Hubble parameter, the symmetry group is not the de Sitter one. However, since this problem does not arise for the ground state in the de Sitter spacetime, one may hope that it  does not also emerge in our setup, since the quantum scalar is assumed to be placed in its ground state. Similarly in \cite{deb2}, certain integrals of the second order metric and matter fluctuations are shown to be much larger than the similar integrals involving first order terms in FRW  spacetimes close to the de Sitter background. Although the main argument of \cite{deb2} is based on purely classical manipulations, it is also proved in the same paper that when fluctuations are taken to be quantum distributions, the result is not spoiled by quantum anomalies. The analysis of \cite{deb2} shows that the strength of the perturbations does not necessarily agree with the order of the perturbative expansion. In our method, we simply ignore the metric perturbations and consider a {\it free} quantum scalar field, i.e the backreaction is not determined in a perturbative expansion, so the complication discussed in \cite{deb2} is expected not to arise in our analysis.

\section{Adding matter}

In the previous section we have seen that the quantized massless
scalar field in a FRW space-time gives positive acceleration. In this section, we include
contributions  coming from other viable cosmological sources and try to see whether
this result  continues  to hold. As usual we assume that these sources, which we call matter, 
can be modeled  by a perfect fluid  obeying the equation of state 
\be
P_M= w  \, \r_M.
\ee
We further assume that  matter does not directly couple to the scalar field and thus its
stress-energy tensor obeys the standard conservation equation, which gives
$\r_M=1/a^{3(1+w)}$. On the other hand, the vacuum energy density continues to 
obey \eq{fc} and the field equations become  
\bea
&& H^2=\fr{8\pi}{3M_p^2}\,(\r_V+\r_M),\label{fm1}\\
&&
\dot{H}+H^2=\left(1+\fr{H^2}{6\pi M_p^2}\right)^{-1}\left[\,\fr{H^4}{3\pi M_p^2}-\fr{
4\pi(1+3w)}{3M_p^2}\r_M\right].\label{fm2} 
\eea
Initially the scalar is placed in its ground state at time $t_0$ which means
\be
\r_V(t_0)=0.
\ee
Typically  the time $t_0$ can be thought to  correspond to big-bang, after which
the largest possible particle creation or vacuum fluctuation effects are
supposed to occur.  

From the decomposition given in \eq{v1} and \eq{v2}, and recalling that
$P_S=-\r_S/3$, one finds $P_V(t_0)\not =0$. This is actually necessary to
satisfy the stress-energy conservation, since   ``the particle creation
effects'' give  $\dot{\r}_V(t_0)>0$ and the conservation equation implies
$P_V(t_0)=-\dot{\r}_V(t_0)/(3H_0)<0$. Therefore at least in the beginning of its evolution the 
vacuum has negative pressure in an expanding universe.   

Using $\r_V(t_0)=0$, one can see from \eq{fm1} and  \eq{fm2} that to obtain
acceleration at $t_0$ the initial expansion speed must obey 
\be \label{ach}
H(t_0)^2>(1+3w)\fr{3\pi}{2}M_p^2.
\ee
Similarly for
\be\label{ih} 
H(t_0)^2>9(1+w)\pi M_p^2,
\ee
one has  $\dot{H}(t_0)>0$. Recall that having acceleration is necessary for the
validity of the the stress-energy tensor  expressions \eq{v1} and \eq{v2}, since
otherwise the modes which were born as  superhorizon and later become subhorizon
demand a special treatment. When the initial expansion speed obeys \eq{ach}, 
the corresponding initial negative pressure of vacuum fluctuations turns out to
be large enough to produce acceleration even when matter has $w>-1/3$. Note that
for $w>-1/3$, \eq{ach} in general gives $H(t_0)\sim
M_p$. One must be aware of the fact that near such Planck scale expansion speeds
the quantum gravitational effects must be taken into account and it is difficult
to trust purely classical equations. Keeping this remark in mind, let us proceed
to analyze the field equations in different cases.  

\subsection{Cosmological constant}

Assume now that in addition to the quantum massless  scalar field, there also exists
a cosmological constant obeying $P_M=-\r_M$ and $\dot{\r}_M=0$.
When the backreaction of quantum fluctuations is ignored  the scale factor becomes
$a=1/(H_0\eta)$, where  $H_0$ is the constant Hubble parameter of the de Sitter
space, and the corresponding mode functions (obeying \eq{mf} and suitable
initial conditions) can be fixed as  
\be
\m_k=\fr{1}{\sqrt{2k}}\left[1-\fr{i}{k\eta}\right]\, e^{-ik\eta}.
\ee  
Using $\m_k$ in \eq{frho} and \eq{fp} one finds that $\r_V=P_V=0$. Therefore, in
the {\it exact} de Sitter background the stress-energy tensor of vacuum
fluctuations vanish as in the case of the flat space.\footnote{This fact has been
used in \cite{mot} to criticize the inflationary mechanism of generating
cosmological perturbations. This criticism was answered in \cite{st} by pointing
out that inflation does not correspond to an exact de Sitter phase.} However,
taking an exact de Sitter space in reality is very questionable  since there
inevitably exists other cosmological sources. Moreover, during inflation, which
is supposedly the era closest to the de Sitter phase,  there exists an "effective" varying 
cosmological constant generated by classical scalar fields. In that case, the
particle creation effects are not  negligible and the vacuum energy density
obeying \eq{fc} necessarily grows.   

To analyze  the evolution let us rewrite \eq{fm2} as 
\be
\dot{H}=\left(1+\fr{H^2}{6\pi M_p^2}\right)^{-1}\left[\,\fr{H^4}{6 \pi M_p^2}-H^2+\fr{
8\pi}{3M_p^2}\r_M\right].\label{fmy} 
\ee
For
$\r_M>9M_p^4/16$ the right hand side is strictly positive for any
value of $H$, which implies $\dot{H}>0$. In that case the Hubble parameter
indefinitely increases and from \eq{fmy} it is possible to see that $H$ blows up
in a finite proper time.  

For $\r_M=9M_p^4/16$, the polynomial in the square brackets in \eq{fmy} can
be written as $[H^2-3\pi M_p^2]^2$. On the other hand, the initial expansion speed
can be determined from \eq{fm1} as $H_0^2=3\pi M_p^2/2$ (recall that we set
$\r_V(t_0)=0$), which is smaller than the double root of the polynomial
$3\pi M_p^2$. Therefore from \eq{fmy},  one sees that starting from its initial
value $H_0^2=3\pi M_p^2/2$, the Hubble parameter increases in time to reach asymptotically
the value   $H^2=3\pi M_p^2$.   

Finally for $0<\r_M<9M_p^4/16$  the polynomial on the right hand side of
\eq{fmy} can be written as 
\be
\fr{1}{6\pi M_p^2}\left[(H^2-H_-^2)(H^2-H_+^2)\right],
\ee
where
\be\label{hpm}
H_{\pm}^2=3\pi M_p^2 \pm3\pi M_p^2 \sqrt{1-\fr{16\r_M}{9M_p^4}}.
\ee
This time the initial expansion speed, which is determined by $\r_M$ as
$H_0^2=8\pi\r_M/(3M_p^2)$, turns out to be smaller than $H_-$. For $H<H_-$, one
sees from \eq{fmy}  that $\dot{H}>0$. Therefore, in that case $H$ will again increase
and become asymptotically $H_-$.   

In summary, we find that for $0<\r_M\leq 9M_p^4/16$ the vacuum fluctuations
of the massless scalar field {\it renormalizes} the bare cosmological constant by
increasing its value from $H_0$ to $H_-$ given in \eq{hpm}. For larger values of
the bare cosmological constant $\r_M>9M_p^4/16$, the scalar field 
yields a big-rip singularity in a finite proper time.  

\subsection{Radiation}

As an important case, let us now consider the quantum scalar field in the
presence of radiation $P_M=\r_M/3$.  From \eq{ach} we see that to get initial
accelerated expansion the Hubble parameter must be very large $H_0\geq
\sqrt{3\pi}M_p$. Thus, as noted above, the results presented in this subsection are likely
to be modified by quantum gravitational effects. In any case, let us proceed 
to see how vacuum fluctuations of the scalar field naively changes the evolution of radiation. 

For $H_0=\sqrt{3\pi}M_p$, the equations \eq{fm1} and \eq{fm2} can be solved
exactly which gives
\be
a=\fr{t}{t_0},
\ee
where $t_0=1/(\sqrt{3\pi}M_p)$. Here the initial expansion speed is tuned to yield
$\ddot{a}=0$, which then implies that both $H^4$ and $\r_M$ in \eq{fm2} 
decrease like $1/a^4$ and cancel each other for all times. Integrating \eq{fc}
the evolution of the vacuum energy density can be found as 
\be
\r_V=\fr{3M_p^2}{8\pi}\left[1-\fr{t_0^2}{t^2}\right]\fr{1}{t^2},\hs{13}t\geq
t_0.
\ee
As pointed out in the previous section nearly in one Hubble time $\r_V$
increases to a maximum value, of the order of $H^4$, and then decreases like
$1/a^2$. 

For $\sqrt{12\pi}M_p>H_0>\sqrt{3\pi}M_p$ one has $\ddot{a}(t_0)>0$ and 
$\dot{H}(t_0)<0$. Since $H^4=\dot{a}^4/a^4$, $\r_M=C/a^4$ and initially
$\dot{a}$ increases (since $\ddot{a}(t_0)>0$), $\r_M$ decreases faster than
$H^4$, which shows by \eq{fm2} that $\ddot{a}$ continues to be positive. On the other hand,
by a similar comparison it is possible to see that
$\dot{H}<0$. Therefore, for this given initial expansion rate the universe
continues to expand with positive acceleration since radiation always decays
faster than the vacuum energy density. As radiation becomes negligible the
evolution can be approximately described by the vacuum only solution \eq{im} (see figure 1-b). 
Note that for $H_0^2<3\pi M_p^2$ our equations cannot be used to analyze the expansion
of the universe since evolution starts with deceleration. 

\subsection{Pressureless Dust}

Finally let us determine the cosmological evolution in the presence of 
pressureless dust with $P_M=0$. From \eq{ach}, for  $H_0>\sqrt{3\pi/2}M_p$ the 
acceleration is initially positive. On the other hand, \eq{ih}
implies that for $H_0\geq 3\sqrt{\pi}M_p$ the initial rate of change of $H$ is
also non-negative. Again, these expansion speeds are very large and the results
presented in this subsection must be taken with care. 

Rewriting \eq{fm2} as
\be
\dot{H}=\left(1+\fr{H^2}{6\pi M_p^2}\right)^{-1}\left[\,\fr{H^4}{6\pi M_p^2}-H^2-\fr{
4\pi}{3M_p^2}\r_M\right].\label{fmd} 
\ee
one sees that for $H_0\geq 3\sqrt{\pi} M_p$, which gives $\dot{H}(t_0)\geq0$, the
first two terms in the square brackets are at least non-decreasing while $\r_M$
decreases. In that case $\dot{H}$ will always be positive and
the dust will become more and more insignificant at later times. Thus, the
metric must approach the purely vacuum solution \eq{im} with increasing $H$ and 
one encounters a big-rip singularity in a
finite proper time (see figure \ref{fig1}-a). 

For $3\sqrt{\pi}M_p>H_0>M_p\sqrt{3\pi/2}$, the expansion starts with acceleration
 but by comparing the positive and negative terms in \eq{fmd} one can see 
 that the  Hubble parameter always decreases.  To determine whether  the acceleration
survives one must compare the rate of change of the two terms in
\eq{fm2}, namely $H^4$ and $\r_M$, which give positive and negative
contributions, respectively. If $H^4$ decreases faster than $\r_M$, the
acceleration will eventually stop. Since $\r_M=C/a^3$, the ratio $H^4/\r_M$ 
becomes proportional to
$f(t)=\dot{a}^4/a$. By using \eq{fm2} one can calculate  
\be
\dot{f}=\dot{a}^3\left[4\fr{\ddot{a}}{a}-\fr{\dot{a}^2}{a^2}\right]\sim 
\left[ \fr{7H^4}{6\pi M_p^2}-H^2-\fr{16\pi\r_M}{3M_p^2} \right],
\ee
which shows that $\dot{f}<0$ for $H^2<6\pi M_p^2/7$. Since $H$ continuously
decreases $\dot{f}$ will eventually become negative and $f$ will  start
decreasing. This proves that in the end $H^4$ term decays faster than $\r_M$
and thus acceleration must stop at some time and deceleration should start
over. Unlike the case with radiation, the acceleration supported by vacuum
fluctuations will be defeated by the deceleration of the pressureless dust. 

\section{Conclusions}

In this paper we try to determine the cosmological evolution of a FRW space-time
driven by the vacuum fluctuations of a quantized massless scalar field. It turns out
that to solve this problem one must deal with a coupled integro-differential
equation system involving the mode functions of the scalar field and the scale
factor of the FRW metric. Namely, the mode functions obey the massless
Klein-Gordon equation on the background and thus they are affected  by the
expansion; at the same time their integral in momentum space determine the
stress-energy tensor of the scalar field, which then fixes the evolution of the
scale factor. To simplify these equations we try to pin down  the stress-energy 
tensor by noting that subhorizon modes evolve nearly adiabatically
and thus their contribution  to any adiabatically regularized expression can be
neglected. On the other hand, since superhorizon modes have equation of state
parameter $\o=-1/3$, their contribution to the stress-energy tensor can be
encoded by one unknown function. One can then use stress-energy conservation to
determine the time evolution of the total vacuum energy density and pressure. In this way
(when the background is accelerating) a simple and consistent set of equations
for two unknown functions, which are the scale factor of the universe and the
vacuum energy density,  can be obtained.  

The above mentioned approximate stress-energy tensor of vacuum fluctuations  
has some peculiar properties. May be the most important  new feature is
that it appears to depend on the Hubble parameter $H$ and its time derivative $\dot{H}$.
The stress-energy conservation yields a non-homogeneous first order
linear equation for the vacuum energy density. The non-homogenous 
source terms are given by  the Hubble parameter and they dictate how vacuum
energy density is forced to increase by the expansion. Similarly, the
homogeneous part of this equation shows how vacuum energy density decreases
with expansion.

We are able to solve the system exactly and obtain analytical expressions
describing the  cosmological evolution of vacuum. From the
field equations one can see that there is always positive acceleration, of the
order of $H^4/M_p^2$. Since the acceleration depends on the speed of
the expansion, the exact time evolution becomes completely different than that
of a perfect fluid with $\o<-1/3$. Namely, for different values of the 
initial Hubble parameter, one encounters different behavior. Since we 
are employing adiabatic regularization, there does not appear any extra
renormalization  scale and not surprisingly the critical expansion speed, which
separates different regimes, turns out to be of the order of the Planck scale. 

Since the magnitude of the acceleration $H^4/M_p^2$ is tiny today one may think
that  vacuum fluctuations cannot have any cosmological impact at the present time.
However, once the vacuum energy density starts to increase in the early universe, in
about one Hubble time it reaches the value $H^4$, which is not necessarily very very small 
compared to the total energy density $H^2M_p^2$.  Furthermore, as pointed out above, 
the vacuum energy density then decreases
like $1/a^2$, which is much slower than radiation. Therefore, even though the
acceleration is wiped out by other sources giving  deceleration, the vacuum energy density can still 
be significant compared to the total energy density of the universe at an earlier 
epoch or even at present. 

In this paper we also try to answer how the quantized massless 
scalar field possibly changes the standard
cosmic evolution of matter in the form of cosmological constant, radiation or
pressureless dust. We show rigorously that vacuum fluctuations
can renormalize the magnitude of the cosmological constant by increasing its
bare value. On the other hand, to obtain acceleration in the presence of
radiation or dust the initial expansion speed must be very large, which is of
the order of Planck scale, and thus these calculations must be taken with care. 
In any case, we show that once vacuum is allowed to produce acceleration in the
beginning, it will dominate over the radiation and always give an accelerated
expansion, but it will eventually be defeated by the pressureless dust and the acceleration
will be taken over by deceleration. As a result we show that vacuum fluctuations
can alter the well known standard cosmological evolutions in a very non-trivial
way. 

It might be interesting to extend these results in different directions. First,
one can work out  the cosmological impact of a quantized {\it massive} field. This is
a harder problem since the mass of the scalar field introduces a new scale to
the problem and the behavior of the mode functions will be more
complicated to determine. Obviously, when the Hubble parameter is much larger than the mass of
the scalar, the analysis will be similar to that of the massless case. However,
when the mass becomes larger than the Hubble parameter all modes start to evolve
adiabatically. Second, it would be interesting to study the impact of the
quantized massless scalar field when the universe decelerates. In that case one
must determine the effects of the modes which were born as superhorizon and
later become subhorizon. The contributions of these modes are expected to be
close to radiation with $\o\sim 1/3$. Of course this prediction must be checked by an
explicit calculation. Third, one wonders if a viable scenario of inflation can
be realized in this setup. We noted earlier that by fine tuning it is possible
to obtain a suitable number of e-foldings by vacuum fluctuations and exit from
inflation is naturally achieved. As shown in \cite{a},   
there inevitably exists (as a matter of fact large) deviations  about the
quantum vacuum expectation values. One may tempt to interpret these deviations as 
the source  of cosmological  perturbations and it would be interesting to determine the 
corresponding spectrum.  

Let us note that in some solutions presented in this paper, the Hubble
parameter increases and this corresponds to an effective equation of state
parameter $\o<-1$. It is known that  when the self interactions are taken into
account in determining the stress-energy tensor of the $\l\phi^4$ scalar in the
de Sitter background, one also finds  $\o<-1$ \cite{kw1,kw2} (it is also possible to obtain $\o<-1$ in the so called classical quintom models, for a review see \cite{quintom}). 
In our case, super-acceleration does not arise from interactions
but it appears due to the vacuum pressure becoming very negative at very large
expansion speeds. This happens since at such speeds the vacuum energy density
tend to increase faster (due to large particle creation effects) and
conservation of the stress-energy tensor necessarily gives correspondingly large
negative pressure. We see here that when super-acceleration occurs due to vacuum 
fluctuations  there always appears a big-rip singularity in a finite proper time. This can be
interpreted as a semi-classical instability of general relativity in the
presence of quantized fields and it would be interesting to sharpen this
observation by other semi-classical methods. 

Finally, in this paper we consider adiabatic subtractions up to second order. In principle, one can repeat the above calculations with the fully regularized stress-energy tensor by including the fourth order adiabatic subtraction terms. As noted above, in that case the field equations contain fourth order time derivatives of the scale factor and to specify the evolution uniquely one must impose not only the initial expansion speed but also the initial acceleration and the third derivative of the scale factor. Although this is puzzling since the physical fields are expected to obey second order dynamical equations, it would be interesting to study the cosmic evolution  with these terms included.

\end{document}